\pdfoutput=1
\documentclass[11pt]{article}
\usepackage{graphicx}
\usepackage{latexsym}
\usepackage{mathrsfs}
\usepackage[scr=rsfs,cal=boondox]{mathalfa}
\usepackage[overload]{textcase}

\font\grande=cmr9.5 scaled \magstep4
\font\medio=cmr9.5 scaled \magstep2
\outer\def\beginsection#1\par{\medbreak\bigskip
      \message{#1}\leftline{\bf#1}\nobreak\medskip
\vskip-\parskip
      \noindent}

\begin{document}
\bibliographystyle{unsrt}

\titlepage
\vspace{1cm}
\begin{center}
{\grande Viscous absorption of ultra-high-frequency gravitons}\\
\vspace{1.5 cm}
Massimo Giovannini \footnote{e-mail address: massimo.giovannini@cern.ch}\\
\vspace{1cm}
{{\sl Department of Physics, CERN, 1211 Geneva 23, Switzerland }}\\
\vspace{0.5cm}
{{\sl INFN, Section of Milan-Bicocca, 20126 Milan, Italy}}
\vspace*{1cm}
\end{center}
\vskip 0.3cm
\centerline{\medio  Abstract}
\vskip 0.5cm
The high-frequency gravitons can be absorbed by the first and second viscosities of the post-inflationary plasma as the corresponding wavelengths reenter the Hubble radius prior to big-bang nucleosynthesis. When the total sound speed of the medium is stiffer than radiation the rate of expansion still exceeds the shear rate but the bulk viscosity is not negligible. Depending on the value of the entropy density at the end of inflation the spectral energy density of the relic gravitons gets modified in comparison with the inviscid result when the frequency ranges between the kHz band and the GHz region. In the nHz
domain the spectrum inherits a known suppression due to neutrino free-streaming but also a marginal spike potentially caused by a sudden outbreak of the bulk viscosity around the quark-hadron phase transition, as suggested by the hadron spectra produced in the collisions of heavy ions. 
\noindent
\vspace{5mm}
\vfill
\newpage
The relic gravitons are produced by the variation of the space-time curvature \cite{FIVE,FIVEa} and their spectral energy density is almost scale-invariant when the conventional inflationary epoch evolves into a radiation-dominated stage of expansion \cite{SIX,SIXa}. Since during radiation the bulk viscosity vanishes \cite{ONE,ONEa} and the shear viscosity is negligible \cite{TWO}, very little absorption is expected at high-frequency. Consequently the quasi-flat plateau (corresponding to the wavelengths that leave the Hubble radius during inflation and reenter in the radiation stage) should remain nearly unmodified except for the free-streaming of the neutrinos \cite{FOUR,SEVEN} and for some possible burst of viscosity, for instance associated with the quark-hadron phase \cite{EIGHT,NINE}. If the post-inflationary expansion rate sharply deviates from radiation for temperatures higher than ${\mathcal O}(0.1)$ MeV, the bulk viscous stresses are not bound to vanish and the ultra-high-frequency gravitons of inflationary origin can be comparatively more absorbed than in the conventional (inviscid) case. This happens, in particular, when the total sound speed $c_{st}$ is stiffer than radiation  (i.e. $c_{st}^2 = w > 1/3$, where $w= p_{t}/\rho_{t}$ denotes the barotropic index). The spectral energy density of the relic gravitons develops then a blue slope and a broad spike \cite{TEN1,TEN2}. A primordial stiff phase (originally suggested in \cite{TENa,TENb}) is realized in quintessential inflationary scenarios \cite{TENc} as well as in a number of similar frameworks (see for instance \cite{TENd,TENe,TENf} for some recent suggestions). The purpose of the present analysis is to demonstrate that the high-frequency gravitons (i.e. between the MHz and the GHz) are absorbed, at different rates, 
depending on their  typical wavelengths, on the entropy density of the fluid and on the total sound speed of the plasma.

After a stage of conventional inflationary expansion the energy-momentum tensor can always be expressed as the sum of an inviscid component supplemented by the viscous contribution\footnote{The signature of the metric is mostly minus (i.e.  $[+,\,-,\,-\,-)$]. The lowercase Greek indices are four-dimensional; the lowercase Latin indices are three-dimensional. The standard four-dimensional projector is defined as ${\mathcal P}_{\beta}^{\,\,\,\nu} = (\delta_{\beta}^{\,\,\,\nu} - u_{\beta} u^{\nu})$ and it is orthogonal to the direction of the four-velocity (i.e. $u_{\nu} \, {\mathcal P}_{\beta}^{\,\,\,\nu}=0$).  The  prefixes of the international system of units will systematically employed (e.g. $1\, \mathrm{nHz} = 10^{-9} $ Hz, $1\, \mathrm{GHz} = 10^{9} $ Hz and so on and so forth for the other typical frequencies mentioned hereunder).}:
\begin{equation}
T_{\mu}^{\,\,\nu} = (p_{t} + \rho_{t}) u_{\mu} u^{\nu} - p_{t} \delta_{\mu}^{\nu} + {\mathcal T}_{\mu}^{\,\,\nu}(\eta,\xi),
\label{EQ1}
\end{equation}
where $u^{\mu}$ is the four-velocity, $p_{t}$ the total pressure and $\rho_{t}$ the total energy density of the plasma. In Eq. (\ref{EQ1}) the viscous contribution is collectively described by ${\mathcal T}_{\mu}^{\nu}(\eta,\xi)$ and it depends on the shear 
and on the bulk viscosities denoted, respectively, by $\eta$ and $\xi$:
\begin{equation}
{\mathcal T}_{\mu}^{\,\,\nu}(\eta,\xi) = 2\, \eta \,\sigma_{\mu}^{\,\,\nu}   + \xi \,\, \theta \,\,{\mathcal P}_{\mu}^{\,\,\nu},\qquad\qquad \theta =\nabla_{\alpha} u^{\alpha}.
\label{EQ2}
\end{equation}
In Eq. (\ref{EQ2}) $\theta$ and $\sigma_{\mu\nu}$ are, respectively, the total expansion and the shear tensor. The projection of the covariant conservation of $T_{\mu}^{\,\,\nu}$ along the two orthogonal directions $u_{\nu}$ and  ${\mathcal P}_{\nu}^{\alpha}$ leads, respectively, to the second principle of thermodynamics and to the relativistic version of the conventional Navier-Stokes equation:
\begin{eqnarray}
&& \nabla_{\mu}[ (p_{t} + \rho_{t}) u^{\mu}] - u_{\alpha} \partial^{\alpha} p_{t} + u_{\beta} \nabla_{\alpha} {\mathcal T}^{\alpha\beta} =0,
\label{EQ3}\\
&& (p_{t} + \rho_{t})  \dot{u}^{\alpha} - \partial^{\alpha}p_{t} + u^{\alpha} u_{\beta} \partial^{\beta} p_{t} + {\mathcal P}^{\alpha}_{\nu} \nabla_{\mu} {\mathcal T}^{\mu\nu} =0,
\label{EQ4}
\end{eqnarray}
where, by definition, $\dot{u}^{\alpha} = u^{\beta} \nabla_{\beta} u^{\alpha}$ is the $4$-acceleration and $\nabla_{\mu}$ denotes throughout the covariant derivative. Using the fundamental thermodynamic identity we obtain, after standard manipulations, the covariant non-conservation of the entropy four-vector:
\begin{equation}
\nabla_{\alpha} [ \mathcal{s} u^{\alpha} - \overline{\mu} \,\,\nu^{\alpha} ] + \nu^{\alpha} \partial_{\alpha} \overline{\mu} = \frac{\nabla_{\alpha} u_{\beta} \, {\mathcal T}^{\alpha\beta}}{T},\qquad\qquad \overline{\mu} = \frac{\mu}{T}.
\label{EQ6}
\end{equation}
where $\mathcal{s}$ denotes the entropy density, $\mu$ is the chemical potential and $T$ is the temperature. 
Equation (\ref{EQ6}) assumes the covariant conservation of the total particle current $j^{\mu} = n_{t}\, u^{\mu} + \nu^{\mu} $ (i.e. $\nabla_{\mu} j^{\mu} = 0$) where $n_{t}$ is the total concentration of the plasma and $\nu^{\mu}$  is the diffusion current. To recover Eq. (\ref{EQ6}) the covariant gradient of the pressure has to be traded for the derivatives of the temperature and of the chemical potential according to the relation $\partial_{\alpha} p_{t} =( \mathcal{s} \partial_{\alpha} T + n_{t} \partial_{\alpha} \mu)$; this relation follows from the first principle of thermodynamics together with the related fundamental identity   $(p_{t} + \rho_{t}) = \mathcal{s} \, T + \mu \, n_{t}$.

The bulk and the shear viscosity have complementary physical features: while the shear viscosity of a relativistic and gravitating fluid is associated with the covariant gradients of the four-velocity, the bulk viscous stresses arise because the expansion of the Universe is continually trying to pull the underlying fluid out of thermal equilibrium. Consequently the shear viscosity never affects the homogeneous and isotropic Friedmannian background while bulk viscous stresses vanish when the equation of state is pure radiation but are otherwise present \cite{ONE,ONEa} (see also \cite{FIFTEEN}). Even though the shear viscosity is the primary source of Silk damping (and affects the compressible modes of the pre-decoupling plasma) its impact on the tensor fluctuations of the geometry is much less relevant. As originally argued by Hawking \cite{TWO} also gravitational waves are absorbed by a viscous medium at a rate $\Gamma_{g} = 2 \ell_{P}^2 \eta$ where, within the present notations, $\ell_{P} = \sqrt{8 \pi G} = 1/\overline{M}_{P}$. This conclusion concerns relativistic plasmas where the collision frequency of the medium exceeds the Hubble rate but it does not apply, for instance, to the case of distant sources \cite{THREE}. When  the collision frequency of the  plasma is smaller than the Hubble rate the hydrodynamic approximation is untenable  and should be replaced by the Boltzmann hierarchy associated with the viscous medium:  this is what ultimately happens in the concordance paradigm 
where the low-frequency gravitational waves are suppressed by the free-streaming of (nearly massless) neutrinos \cite{FOUR}.  The explicit form of Eq. (\ref{EQ6}) has been obtained by trading the term $u_{\nu} \nabla_{\mu} {\mathcal T}^{\mu\nu}$ for $(\nabla_{\mu} u_{\nu}) {\mathcal T}^{\mu\nu}$ since, in the Landau frame, $\nabla_{\mu} ( u_{\nu} {\mathcal T}^{\mu\nu} ) = 0$. We remind that the viscous energy-momentum tensor can be evaluated either in the Landau-Lifshitz or in the Eckart frames\footnote{In the Eckart case the four-velocity $u_{\mu}$ introduced in Eq. (\ref{EQ1}) denotes the velocity of the particle transport (see e.g. \cite{FIFTEEN}). The Eckart frame is fixed by requiring that $j^{\alpha} \, u_{\alpha} =0$ while ${\mathcal T}^{\mu\nu} u_{\nu} \neq 0$.  Conversely, in the Landau-Lifshitz approach (which is the one adopted here) pure thermal conduction corresponds to an energy flux without particles: the four-velocity $u_{\mu}$ coincides with the velocity of the energy transport implying ${\mathcal T}^{\mu\nu} u_{\nu} =0$.}.  As a consequence, in agreement with  ${\mathcal T}^{\mu\nu} u_{\nu} =0$, the explicit expression of the shear tensor implies $u_{\mu} \,\sigma^{\mu\nu} =0$:
\begin{equation}
\sigma_{\mu\nu} = {\mathcal P}_{\mu}^{\,\,\alpha} {\mathcal P}_{\nu}^{\,\,\beta} W_{\alpha\beta}, \qquad \qquad 
W_{\alpha\beta} = \nabla_{\alpha} u_{\beta} + \nabla_{\beta} u_{\alpha} - \dot{u}_{\alpha} u_{\beta} - \dot{u}_{\beta} u_{\alpha} - \frac{2}{3} g_{\alpha\beta} \theta,
\label{EQ7} 
\end{equation}
where, as usual,  $\theta = \nabla_{\lambda} u^{\lambda}$ is the expansion scalar. Therefore, using Eq. (\ref{EQ7}) and neglecting the chemical potential the expression of the second principle given in Eq. (\ref{EQ6}) is:
\begin{equation}
\nabla_{\mu} \mathcal{s}^{\mu} = \xi\, \,\frac{\theta^2}{T} + \eta \,\,\frac{\sigma^2}{T},\qquad\qquad \mathcal{s}^{\mu} = \mathcal{s} \, u^{\mu}, \qquad \qquad \sigma^2 = \sigma_{\mu\nu} \sigma^{\mu\nu}.
\label{EQ8}
\end{equation}
In a conformally flat background metric $\overline{g}_{\alpha\beta} = a^2(\tau)\, \eta_{\alpha\beta}$ (where $a(\tau)$ is the scale factor and $\tau$ denotes the conformal time coordinate) the contribution of the shear vanishes (i.e. $\sigma^2 =0$) so that Eq. (\ref{EQ8}) becomes:
\begin{equation}
\mathcal{s}^{\prime} + 3 {\mathcal H} \mathcal{s} = \frac{9 \, \,\xi\, {\mathcal H}^2}{a (p_{t} + \rho_{t})} \,\mathcal{s},\qquad {\mathcal H} = \frac{a^{\prime}}{a}, \qquad \theta= 3 \frac{{\mathcal H}}{a},
\label{EQ9}
\end{equation}
where the prime denotes throughout a derivation with respect to $\tau$; in this respect
the relation between ${\mathcal H}$ and the standard Hubble rate is given, as usual, by $a\, H= {\mathcal H}$.
It is finally relevant to appreciate, for immediate convenience, that  at the right-hand side of  Eq. (\ref{EQ9}) the temperature has been eliminated by using the fundamental identity of thermodynamics in the absence of chemical potential [i.e. $T\, \mathcal{s} = (p_{t} + \rho_{t})$].

The absorption of the relic gravitons by the viscosities of the post-inflationary medium is estimated by combining  Eqs. (\ref{EQ3})--(\ref{EQ4}) with the homogeneous and inhomogeneous versions of Einstein's equations whose covariant form is usefully written as:
\begin{equation}
R_{\mu}^{\,\,\,\,\nu} = \ell_{P}^2 \biggl(T_{\mu}^{\,\,\,\,\nu} - \frac{T}{2} \delta_{\mu}^{\,\,\,\,\nu}\biggr), \qquad\qquad T= T_{\mu}^{\,\,\,\mu},
\label{EQ11}
\end{equation}
where $R_{\mu}^{\,\,\,\,\nu}$ is the Ricci tensor with mixed indices and $T_{\mu}^{\,\,\,\,\nu}$ has been already introduced in Eq. (\ref{EQ1}).
Since the shear tensor of Eq. (\ref{EQ7}) vanishes in a Friedmannian background, the homogeneous evolution equations deduced from Eqs. (\ref{EQ1}) and (\ref{EQ11}) are only affected by the bulk viscosity and they are:
\begin{equation}
2({\mathcal H}^2 - {\mathcal H}^{\prime}) = \ell_{P}^2 \, a^2(\rho_{t} + p_{t}) - 3\ell_{P}^2 \, \,a\,\, {\mathcal H}\, \, \xi, \qquad\qquad 3{\mathcal H}^2 = \ell_{P}^2 \, a^2 \rho_{t}.
\label{EQ12}
\end{equation}
The homogeneous version of Eq. (\ref{EQ3})  leads instead to the following expression:
\begin{equation}
\rho_{t}^{\prime} + 3 {\mathcal H} (\rho_{t} + p_{t}) = \frac{9\,\,\xi\,\, {\mathcal H}^2}{a}.
\label{EQ13}
\end{equation}
If the source term at the right-hand side of Eq. (\ref{EQ13}) is eliminated with the help of Eq. (\ref{EQ9}), the following explicit  relation
\begin{equation}
\frac{\rho_{t}^{\prime}}{\rho_{t} + p_{t}} = \frac{\mathcal{s}^{\prime}}{\mathcal{s}}\qquad \Rightarrow \qquad
\frac{d \, \rho_{t}}{p_{t} + \rho_{t}} = d\ln{\mathcal{s}},
\label{EQ13a}
\end{equation}
 connects the energy and the entropy densities to the total pressure.
While Eqs. (\ref{EQ12})--(\ref{EQ13}) and (\ref{EQ13a}) are not affected by the value of $\eta$, the opposite is true for the inhomogeneities since the first-order (tensor) fluctuation of the shear viscosity does not vanish and it easily obtained from Eq. (\ref{EQ7}):
\begin{equation}
\delta_{t}^{(1)} g_{i j} = - a^2 h_{i j}, \qquad \delta_{t}^{(1)} \sigma_{i j}  = - a h_{ij}^{\prime}, \qquad \delta_{t}^{(1)} \, \sigma_{i}^{\,\,\,j} = h_{i}^{\,\,j\, \prime}/a, \qquad\qquad \partial_{i} h^{i}_{j} =h_{i}^{i} =0,
\label{EQ13b}
\end{equation}
where $\delta_{t}^{(1)}$ denotes the first-order (tensor) fluctuation of the corresponding 
quantity. The evolution of $h_{ij}$ follows from Eq. (\ref{EQ11}) when the fluctuations of the 
Ricci tensor are combined with the ones of the viscous sources so that 
the result (already anticipated above and originally derived in Ref. \cite{TWO}) is:\footnote{The propagation of gravitational waves in curved backgrounds can be notoriously treated either within the covariant approach (analog to the one pioneered by Lifshitz \cite{TENca} and employed in Ref. \cite{TWO}) or in a non-covariant description which is particularly suitable for cosmological applications. In the covariant approach the 
tensor mode is defined as $u_{\mu} f^{\mu}_{\nu} = \overline{\nabla}_{\mu} f^{\mu}_{\nu} = \overline{g}^{\mu\nu} f_{\mu\nu} =0$ where $\overline{g}_{\mu\nu}$ denotes the background metric and $f_{\mu\nu}$ is the corresponding fluctuation; $\overline{\nabla}_{\mu}$ is the covariant derivative defined with respect to $\overline{g}_{\mu\nu}$. These two complementary approaches are fully equivalent and have been recently reviewed (at length) in Ref. \cite{TENcb}.}:
\begin{equation}
h_{i}^{\,j\,\prime\prime} + (2 {\mathcal H}  + \Gamma_{g} a) h_{i}^{\,j\,\prime} - \nabla^2 h_{i}^{\,j} =0, \qquad\qquad \Gamma_{g} = 2 \ell_{P}^2 \eta.
\label{EQ14}
\end{equation}
While Eq. (\ref{EQ14}) describes the tensor inhomogeneities after the end of inflation,
in the past $\xi$ has been suggested as the driving source of quasi-de Sitter stage of expansion. If this is the case the scalar fluctuations of $\xi$ also dominate the scalar inhomogeneities by inducing a quasi-adiabatic solution which is however strongly suppressed; this mode cannot be a substitute for the conventional adiabatic paradigm since it  leads to an anomalously large tensor to scalar ratio \cite{TENg}. While the bulk viscosity cannot drive a conventional inflationary stage of expansion,  $\xi$ may also affect the character of the cosmological singularity \cite{ELEVEN,TWELVE};  in this context the viscosities may lead to curvature bounces where the spectral energy density of the relic gravitons has a blue slope \cite{THIRTEEN,FOURTEEN}. Viscosities are not regarded here as speculative early-time sources potentially unrelated to the thermodynamical properties of the medium but rather as a specific physical features of the late-time (post-inflationary) evolution. 
 
In a relativistic plasma the total 
energy density $\rho_{t}$ (already introduced in Eq. (\ref{EQ1})) and the collision frequency $\Gamma_{coll}$ determine the corresponding viscosities \cite{ONE,ONEa,FIFTEEN}:
\begin{equation} 
\eta= b_{\eta} \frac{\rho_{t}}{\Gamma_{coll}}, \qquad \qquad \xi = b_{\xi} \frac{\rho_{t}}{\Gamma_{coll}}
\biggl(c_{st}^2 - \frac{1}{3}\biggr)^2, 
\label{EQ2a}
\end{equation}
where $b_{\eta}$ and $b_{\xi}$ are two (dimensionless) numerical constants while, as anticipated, $c_{st}^2 =p_{t}^{\prime}/\rho_{t}^{\prime}$ denotes the total sound speed of the plasma. Based on Eq. (\ref{EQ2a})  $\Gamma_{g}$ and $\Gamma_{coll}$ are inversely proportional in a radiation-dominated stage where the shear viscosity solely depends on the entropy density and the bulk viscosity vanishes:
\begin{equation}
\frac{\Gamma_{g}}{H} \propto \frac{H}{\Gamma_{coll}}, \qquad\qquad \eta \propto \mathcal{s}, \qquad\qquad \frac{\xi}{\eta} \to 0,
\label{EQ2b}
\end{equation}
where, as usual, $H = {\mathcal H}/a$ is the standard Hubble rate. As long as $\Gamma_{coll} \gg H$,  the first relation in Eq. (\ref{EQ2b}) explains why $\Gamma_{g}$  can be neglected in Eq. (\ref{EQ14}) and the viscosities have a little impact on the gravitational wave propagation during a radiation-dominated stage. When $\Gamma_{coll} < H$  the hydrodynamic approximation fails and we must 
resort to the full Einstein-Boltzmann hierarchy as in the case of the damping induced 
by the free-streaming of neutrinos \cite{FOUR} at low-frequencies ${\mathcal O}(\mathrm{nHz})$. Equation (\ref{EQ2b}) also describes the more general situation of a relativistic plasma where  the particles  interact by  the exchange of gauge bosons (like in the case of a generalized Coulomb plasma). Neglecting the Coulomb logs, 
 the collision frequency is $\Gamma_{coll} \simeq  \alpha^2 T$ while $\rho_{t} = (\pi^2/30) {\mathcal N} T^4$ where ${\mathcal N}$ denotes the effective numbers of relativistic degrees of freedom.  As a consequence Eq. (\ref{EQ2a}) implies that $\xi \to 0$ while $\eta \propto {\mathcal N} T^3/\alpha^2$. Recalling now Eq. (\ref{EQ14}) the shear rate, as expected, will always be $\Gamma_{g} \ll H$ whenever $T < \alpha^2 \overline{M}_{P}$. Since the cross section for the interaction of two gravitons is given by 
$\sigma_{g} = \ell_{P}^2 (T^2/\overline{M}_{P}^2)$  the 
interaction rate $\Gamma_{int} \simeq \sigma_{g} T^3$ is smaller than the Hubble rate 
provided $T < \overline{M}_{P}$. All in all among $\Gamma_{coll}$, $\Gamma_{g}$ and $\Gamma_{int}$ the following hierarchies exist:  
\begin{equation} 
\eta = \overline{b}_{\eta} \mathcal{s}, \qquad \Gamma_{int} < H, \qquad \Gamma_{g} < H, \qquad \Gamma_{coll} > H,
\label{EQ2c}
\end{equation}
where $\overline{b}_{\eta} = b_{\eta}/\alpha^2$ and $\mathcal{s} \propto {\mathcal N}\,T^3$. When discussing the the viscosity bounds\footnote{It is often argued that in all gauge theories with Einstein gravity duals $4 \pi \eta/\mathcal{s} \geq 1$ so that strongly coupled QCD plasma may even saturate 
this bound \cite{FOURTEENa}. }  $\xi$ and $\eta$ are measured in units of $\mathcal{s}$ (see e.g. \cite{NINE}) and the same strategy will be adopted here.

Let us now move to the physically interesting situation where the inflationary stage is followed by a stiff epoch; for the sake of illustration
it is useful to focus on the case $c_{st}^2\to 1$ implying, in the present units, that the total sound speed coincides with the speed of light \cite{TEN1,TEN2,TENa,TENb,TENc} (see also \cite{TENd,TENe,TENf}). According to Eq. (\ref{EQ2a}) both viscosities 
are present in the plasma and they can be parametrized as:
\begin{equation}
\eta = \overline{b}_{\eta} \mathcal{s}, \qquad\qquad \xi = \frac{4\overline{b}_{\xi}}{9} \mathcal{s}, \qquad \qquad
\frac{\eta}{\xi} = \frac{9 \, \overline{b}_{\eta}}{4 \, \overline{b}_{\xi}},
\label{EQ2d}
\end{equation}
where the factor $4/9$  comes from the term $(w-1/3)^2$ in the case $c_{st}^2 = w = 1$. As clarified in Eqs. (\ref{EQ2a})--(\ref{EQ2b}) and (\ref{EQ2c}) it is always possible to parametrize $\xi$ and $\eta$ in terms of the entropy density
however the dependence of $\mathcal{s}$ on $\rho_{t}$ will ultimately follow from Eq. (\ref{EQ13a}). In the case of a stiff plasma where the sound speed 
coincides with $c_{st}=1$ Eq. (\ref{EQ13a}) demands 
that $\mathcal{s} \propto \rho_{t}^{1/2}$ (since $p_{t} = \rho_{t}$); furthermore thanks to Eq. (\ref{EQ2d}) we have
\begin{equation}
\mathcal{s} = \mathcal{s}_{1} \, \biggl(\frac{\rho_{t}}{\rho_{1}}\biggr)^{1/2}, \qquad \qquad 
\eta = \overline{b}_{\eta}\, \mathcal{s}_{1} \, \biggl(\frac{\rho_{t}}{\rho_{1}}\biggr)^{1/2},
\qquad\qquad  \xi = \frac{4\overline{b}_{\xi}\, \mathcal{s}_{1} }{9}  \biggl(\frac{\rho_{t}}{\rho_{1}}\biggr)^{1/2},
\label{EQ2e}
\end{equation}
where  $\mathcal{s}_{1}$ and $\rho_{1}$ denote, respectively, the entropy and the energy densities at the end of inflation while $\overline{b}_{\eta}$ and $\overline{b}_{\xi}$ are two numerical factors both ${\mathcal O}(1)$. If we now insert Eq. (\ref{EQ12}) into Eq. (\ref{EQ13})  and take into account 
the result of Eq. (\ref{EQ2e}) the viscous correction to the evolution of the stiff background are determined from the following equation:
\begin{equation}
\frac{d \mathcal{G}}{d z} + 6 \,\mathcal{G}\, =\, \frac{4}{3} \,\delta\,\, \mathcal{G},
\qquad\qquad \delta = \frac{\overline{b}_{\xi} \,\,\mathcal{s}_{1}}{ H_{1} \overline{M}_{P}^2}, \qquad\qquad \mathcal{G}= \frac{\rho_{t}}{\rho_{1}}< 1,
\label{EQ18}
\end{equation}
where $z = \ln{a}$. The solution of Eq. (\ref{EQ18}) must then be inserted back into Eq. (\ref{EQ12}) with the purpose of determining the explicit form of the scale factor. Since the regularity of the extrinsic curvature of the background demands the continuity of $a \, {\mathcal H}$ across the inflationary boundary,  the 
evolution of the scale factor corrected by the viscosity effects reads:
 \begin{equation}
a_{v}(\tau) = \biggl[ \frac{\beta}{\gamma} \biggl( \frac{\tau}{\tau_{1}} +1 \biggr) +1\biggr]^{\gamma}, \qquad\qquad \tau \geq -\tau_{1},\qquad\qquad \gamma = \frac{3 }{6- 2\delta},
\label{EQ20}
\end{equation}
where the subscript refers to the viscous stage. In Eq. (\ref{EQ20}) 
$\beta = 1/(1-\epsilon)$ and $\epsilon = - \dot{H}/H^2$ is the standard slow-roll parameter defined 
during the inflationary evolution.  Equation (\ref{EQ20}) and its first time derivatives 
are both continuous for $\tau = - \tau_{1}$, i.e. $a_{i}(-\tau_{1}) = a_{v}(-\tau_{1})$ and 
$a_{i}^{\prime}(-\tau_{1}) = a_{v}^{\prime}(-\tau_{1})$, where $a_{i}(\tau)$ denotes the 
inflationary scale factor. The relevant physical situation is the one where ${\mathcal G} <1$, the energy density decreases 
and the scale factor expands; according to Eq. (\ref{EQ20}) this happens for $\delta < 9/2$ even if, in practice,  
this limit is never reached since the physical range of $\delta$ is always below $1$ and  it ultimately depends on the upper bound on $\mathcal{s}_{1}$.

A lower bound on $\mathcal{s}_{1}$ is, approximately, ${\mathcal N} H_{1}^3$ (where ${\mathcal N}$ represents the total number of species). This estimate follows from the particle production at the end of inflation when the post-inflationary expansion rate is slower than radiation, as originally suggested by Ford
\cite{SEVENTEEN} (see also \cite{TEN1}). Since $(H_{1}/\overline{M}_{P})= {\mathcal O}(10^{-6})$ and ${\mathcal N} = {\mathcal O}(100)$ we can argue that, at least, $ \delta \geq \overline{b}_{\xi} {\mathcal N} (H_{1}/\overline{M}_{P})^2 \simeq 10^{-6}$. An upper bound on $\delta$ is instead obtained by assuming that all the energy density at the end of inflation is suddenly transformed in thermal energy density; in actual examples (mostly based on explosive 
 pre-heating) this limit is never reached and it would imply that $T_{1} \sim (H_{1} \overline{M}_{P})^{1/2}$. In this second case  $\mathcal{s}_{1} = {\mathcal O}({\mathcal N} T_{1}^3)$ so that $\delta \leq \overline{b}_{\xi} {\mathcal N} \, (H_{1}/\overline{M}_{P})^{1/2}$; this means that, at most, $ \delta \ll 1$. Since $10^{-6} < \delta \ll  1$ the background evolution 
 is slightly modified by the viscosity, and, in particular the energy density of the produced particles 
will first thermalize and then eventually dominate the background:
\begin{equation}
\biggl(\frac{a_{1}}{a_{\ast}}\biggr) \simeq {\mathcal N}^{3/(6-4\delta)} \, \biggl(\frac{H_{1}}{\overline{M}_{P}}\biggr)^{3/(3 - 2\delta)}, \qquad \qquad \biggl(\frac{a_{1}}{a_{th}}\biggr) \simeq (\alpha^2 {\mathcal N})^{3/(6 - 2 \delta)},\qquad\qquad 10^{-6} \leq \delta \ll 1,
\label{EQ20a}
\end{equation}
where $a_{\ast}$ and $a_{th}$ are, respectively, the scale factors when radiation 
dominates and when the produced fluctuations thermalize\footnote{Equation (\ref{EQ20a}) follows since 
the particles are produced with typical energy density ${\mathcal N} H_{1}^4$ and 
with approximate concentration given by $\mathcal{n} = {\mathcal N} T^3$, where now $T \simeq H_{1}(a_{1}/a)$ 
is the kinetic temperature that eventually coincides with the thermodynamic temperature after thermalization \cite{SEVENTEEN}. If the interactions occur  via the exchange of gauge bosons the cross section is proportional to $\sigma \simeq \alpha^2/T^2$ the thermalization time is defined as $\sigma \mathcal{n} \sim H$; this observation 
 leads to the second result of Eq. (\ref{EQ20a}).}. Since the shift induced by $\delta$ 
is visibly quite small it can be neglected, in the first approximation, when discussing the 
estimates of $a_{th}$ and $a_{\ast}$.

The viscous corrections on the spectral energy density of the relic gravitons 
simply follow by evaluating the impact of $\eta$ and $\xi$ on the evolution 
of the corresponding field operators:
\begin{equation}
\widehat{h}_{i\, j}(\vec{x},\tau) = \frac{\sqrt{2} \, \ell_{P}}{(2\pi)^{3/2}} \sum_{\lambda = \oplus, \, \otimes} \int d^{3} k \,\, e_{i j}^{(\lambda)}(\hat{k})\biggl[ F_{k, \lambda} \,\,\hat{a}_{\vec{k}, \lambda} \,\,e^{- i \vec{k} \cdot\vec{x}} + F^{\ast}_{k, \lambda} \,\,\hat{a}_{\vec{k}, \lambda}^{\dagger} \,\,e^{ i \vec{k} \cdot\vec{x}} \biggr],
\label{EQ21}
\end{equation}
where $e_{i j}^{(\lambda)}(\hat{k})$ denote the two tensor polarizations and $F_{k, \lambda}$ is the mode function. As usual we define the two tensor polarizations 
as $e_{i j}^{\oplus}(\hat{k}) = (\hat{m}_{i} \, \hat{m}_{j} - \hat{n}_{i} \, \hat{n}_{j})$ and as $e_{i j}^{\otimes}(\hat{k}) = (\hat{m}_{i} \, \hat{n}_{j} +\hat{n}_{i} \, \hat{m}_{j})$ where $\hat{m}$, $\hat{n}$ and $\hat{k}$ form a triplet of mutually orthogonal unit vectors. For $\tau \leq - \tau_{1}$ 
the mode functions obey the same evolution for each of the two polarizations so that the index 
$\lambda$ can be dropped:
\begin{equation}
f_{k}^{\prime\prime} + \biggl[ k^2 - \frac{2 -\epsilon}{\tau^2(1- \epsilon)^2}\biggr] f_{k}=0, \qquad \qquad f_{k} = a \, F_{k},\qquad\qquad g_{k} = f_{k}^{\prime} - {\mathcal H} f_{k}.
\label{EQ22}
\end{equation}
In Eq. (\ref{EQ22}) $g_{k}$ denotes the 
 mode function associated with the canonical momentum conjugate to $\widehat{h}_{ij}$.
The solution of Eq. (\ref{EQ22}) with the appropriate boundary conditions for $\tau \to - \infty$ is:
\begin{eqnarray}
f_{k}(\tau) = \frac{N_{\mu}}{\sqrt{2 k}} \, \sqrt{- k\tau} \, H_{\mu}^{(1)}(-k\tau), \qquad
g_{k}(\tau) = - N_{\mu} \,\sqrt{\frac{k}{2}} \,  \sqrt{- k\tau} \, H_{\mu-1}^{(1)}(-k\tau), \qquad \mu= \frac{3 - \epsilon}{2(1 - \epsilon)},
\label{EQ24}
\end{eqnarray}
where $H_{\mu}^{(1)}(-k\tau)$ is the Hankel function of the first kind \cite{EIGHTEEN} and the normalization factor is $N_{\mu} = \sqrt{\pi/2}\,\,e^{i \pi (\mu +1/2)}$.  For $\tau\geq - \tau_{1}$ the tensor amplitude obeys instead Eq. (\ref{EQ14}) where the scale factor 
is given by Eq. (\ref{EQ21}) and the mode function obeys:
\begin{equation}
F_{k}^{\prime\prime} + ( 2 {\mathcal H} + \Gamma_{g} a) F_{k}^{\prime} + k^2 F_{k}=0. 
\label{EQ25}
\end{equation}
Even in the case $c_{st}= \sqrt{w} =1$ the shear rate can be neglected in comparison with the Hubble rate.
In fact, according to Eq. (\ref{EQ2d}), $\eta \propto \mathcal{s}$. Furthermore Eq. (\ref{EQ13a}) demands  $\eta \propto \mathcal{s} \propto \mathcal{s}_{1} \,(\rho/\rho_{1})^{1/2}$ so that the shear viscosity scales exactly as $H$ and it does not affect 
Eq. (\ref{EQ25}) as long as $\delta < 1$:
\begin{equation}
\frac{\Gamma_{g}}{a {\mathcal H}} = \frac{ 2 \,\,\overline{b}_{\eta}\,\, \ell_{P}^2\,\, \mathcal {s}}{H} = 2 \,\,\overline{b}_{\eta}\,\, \delta \ll  1.
\label{EQ25a}
\end{equation}
Under the conditions established by Eq. (\ref{EQ25a}) the mode functions for $\tau \geq - \tau_{1}$ are uniquely determined from Eq. (\ref{EQ24}) in terms of  $\overline{f}_{k}= f_{k}(-\tau_{1})$ and $\overline{g}_{k}= g_{k}(-\tau_{1})$:
\begin{equation}
\left(\matrix{ f_{k}(\tau) &\cr
g_{k}(\tau)/k&\cr}\right) = \left(\matrix{ A_{f\, f}(k, \tau, \tau_{1})
& A_{f\,g}(k,\tau, \tau_{1})&\cr
A_{g\,f}(k,\tau, \tau_{1}) &A_{g\,g}(k,\tau, \tau_{1})&\cr}\right) \left(\matrix{ \overline{f}_{k} &\cr
\overline{g}_{k}/k&\cr}\right),
\label{EQ28}
\end{equation}
where the various entries of the matrix in Eq. (\ref{EQ28}) can be explicitly 
computed in terms of Bessel functions of index $\nu$:
\begin{eqnarray}
A_{f\, f}(k, \tau, \tau_{1}) &=& \frac{\pi}{2} \sqrt{q x_{1}} \sqrt{ k y} \biggl[ J_{\nu+1}( q x_{1}) Y_{\nu}(k y) - Y_{\nu+1}(q x_{1}) J_{\nu}(k y) \biggr],
\nonumber\\
A_{f\, g}(k, \tau, \tau_{1}) &=& \frac{\pi}{2} \sqrt{q x_{1}} \sqrt{ k y} \biggl[ J_{\nu}( q x_{1}) Y_{\nu}(k y) - Y_{\nu}(q x_{1}) J_{\nu}(k y) \biggr],
\nonumber\\
A_{g\, f}(k, \tau, \tau_{1}) &=& \frac{\pi}{2} \sqrt{q x_{1}} \sqrt{ k y} \biggl[ Y_{\nu+1}( q x_{1}) J_{\nu +1}(k y) - J_{\nu+1}(q x_{1}) Y_{\nu+1}(k y) \biggr],
\nonumber\\
A_{f\, g}(k, \tau, \tau_{1}) &=& \frac{\pi}{2} \sqrt{q x_{1}} \sqrt{ k y} \biggl[ Y_{\nu}( q x_{1}) J_{\nu+1}(k y) - Y_{\nu+1}(k y) J_{\nu}(q x_1) \biggr],
\label{EQ29}
\end{eqnarray}
where, thanks to the Wronskians of the Bessel function \cite{EIGHTEEN}, $(A_{f\,f} A_{g\,g} - A_{f\,g} A_{g\,g})=1$ so that 
the matrix of Eq. (\ref{EQ28}) is unitary.
The variables $y=y(\tau, q)$, $q= q(\epsilon,\delta)$ and $\nu= \nu(\delta)$ that appear in Eq. (\ref{EQ29}) are defined, respectively, as:
\begin{equation}
y = y(\tau,q) = \tau + \tau_{1} \biggl( 1 + q\biggr), \qquad
q =q(\epsilon,\delta)= \frac{3(1-\epsilon)}{6 - 2 \delta}, \qquad  \nu = \nu(\delta) = \frac{\delta}{6 - 2 \delta},
\end{equation}
where the dependence on the various arguments has been explicitly indicated even if it will be dropped hereunder to maintain a concise notation. There is in fact a hierarchy between the different contributions appearing in 
Eq. (\ref{EQ29}); in particular it can be easily shown that:
\begin{eqnarray} 
\bigl| A_{f\, f}(k,\tau, \tau_{1}) \, \overline{f}_{k} \bigr| \gg \biggl| A_{f\, g}(k,\tau, \tau_{1}) \frac{\overline{g}_{k}}{k} \biggr|, \qquad\qquad \bigl| A_{g\, f}(k,\tau, \tau_{1}) \, k \, \overline{f}_{k} \bigr| \gg \bigl| A_{g\, g}(k,\tau, \tau_{1}) \overline{g}_{k}\bigr|.
\label{EQ30}
\end{eqnarray} 

The approximation of Eq. (\ref{EQ30}) holds when $x_{1} = k\tau_{1} \ll 1$ and it always verified 
for the estimate of the spectral energy density after the various wavelengths reenter the Hubble radius.
For instance the spectral energy density corresponding to the modes reentering 
during the stiff epoch is given by:
\begin{figure}[!ht]
\centering
\includegraphics[height=7cm]{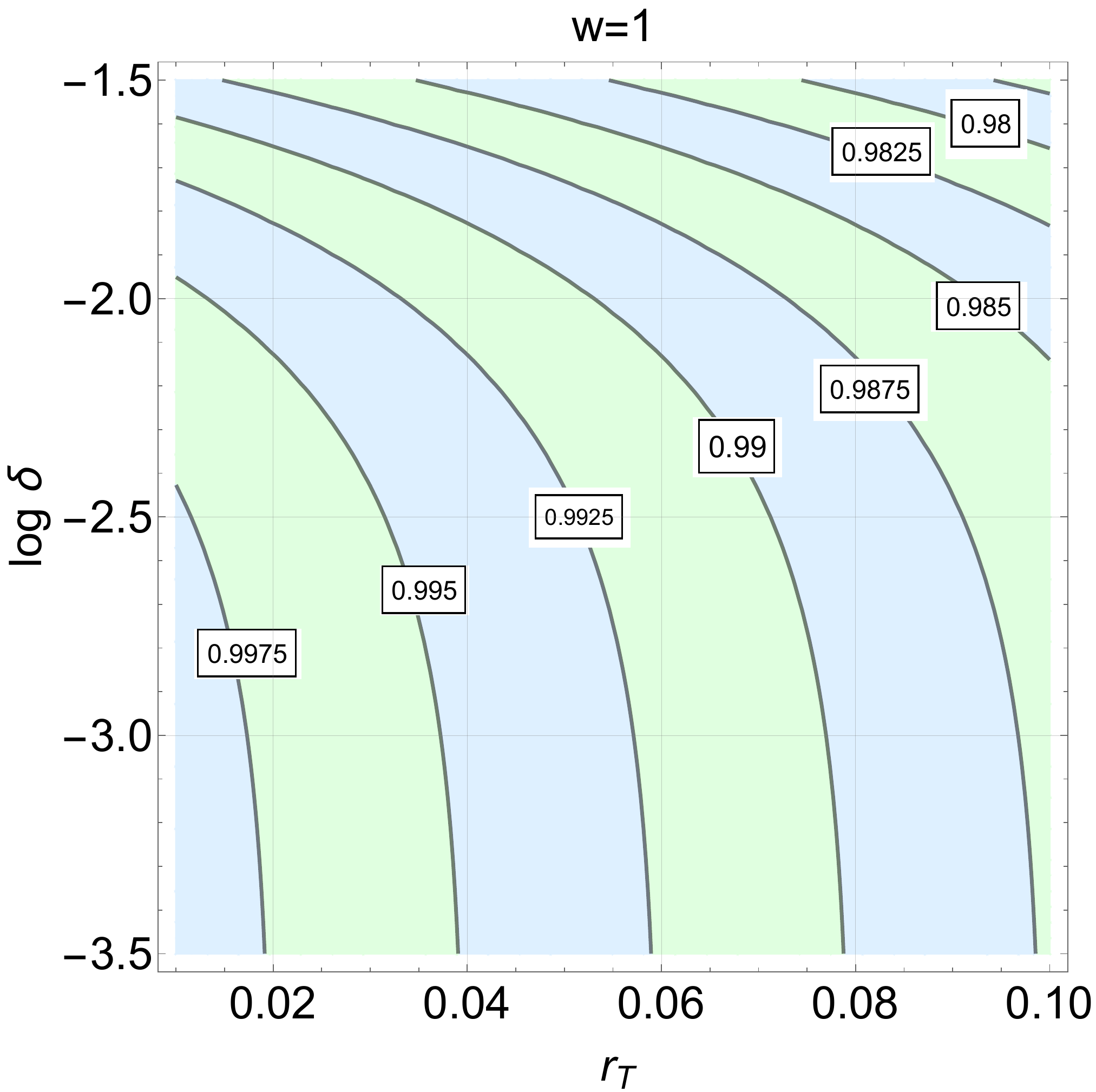}
\includegraphics[height=7cm]{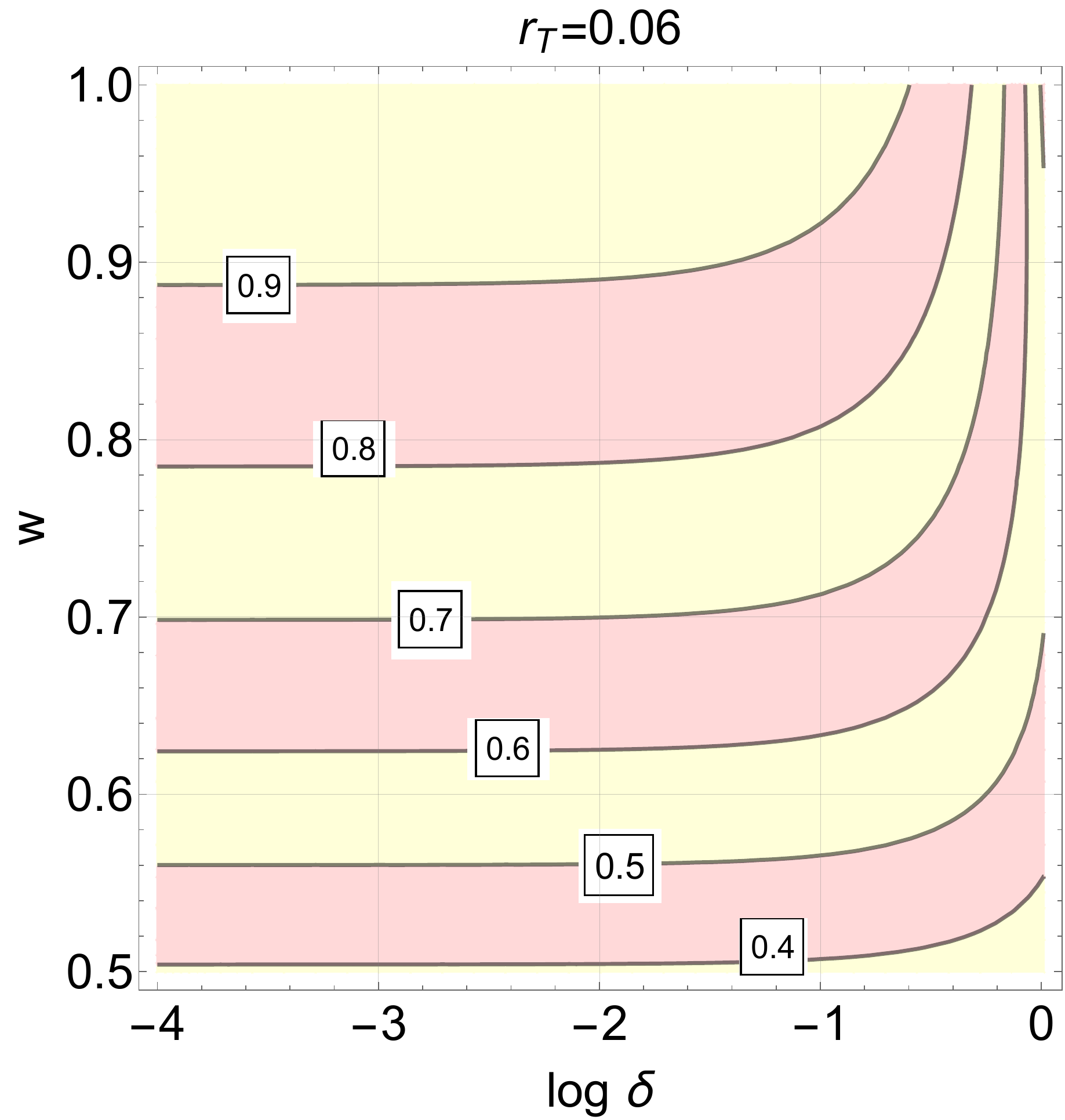}
\caption[a]{The high-frequency spectral index $n_{T}$ reported in Eq. (\ref{EQ32}) is illustrated for different values of $\epsilon= r_{T}/16$ (the slow-roll parameter), $w$ (the barotropic index of the post-inflationary stage) and $\delta$ (the entropy density at the end of inflation and in units of $H_{1} \overline{M}_{P}^2$).}
\label{FIG1}      
\end{figure}
\begin{equation}
 \Omega_{gw}(k,\tau) =  \frac{k^5 \, \bigl|\overline{f}_{k}\bigr|^2}{6 \pi^2 \, H^2 \, a^4\, \overline{M}_{P}^2} \biggl[ \bigl|A_{f\,f}(k, \tau, \tau_{1}) \bigr|^2 +  \bigl|A_{f\,g}(k, \tau, \tau_{1}) \bigr|^2 \biggr].
\label{EQ31}
\end{equation} 
By now inserting Eqs. (\ref{EQ29})--(\ref{EQ30}) into Eq. (\ref{EQ31})  the final expression of the spectral energy density becomes:
\begin{eqnarray}
\Omega_{gw}(k,\tau) &=& {\mathcal C}(\delta, \epsilon) \biggl(\frac{H_{1}}{M_{P}}\biggr)^2 \biggl(\frac{a_{1} H_{1}^2}{a\, H^2}\biggr)^2 \biggl( \frac{k}{a_{1} \, H_{1}}\biggr)^{n_{T}},
\nonumber\\
{\mathcal C}(\delta, \epsilon) &=&\frac{2^{(3- \epsilon)/[2 (1- \epsilon)]} \, 2^{3/(3 - 2\delta)}}{3 \pi^3}
 \,\Gamma^2\biggl[\frac{6 - \delta}{2( 3 - \delta)}\biggr] \, \Gamma^2\biggl(\frac{3 - \epsilon}{2 - 2\epsilon}\biggr)   \biggl[\frac{3(1- \epsilon)}{4(3 - \delta)}\biggr]^{- 2 \delta},
 \label{EQ31a}
 \end{eqnarray}
 where we restored $M_{P}$ by recalling its relation with the reduced Planck mass $\overline{M}_{P} = M_{P}/\sqrt{8 \, \pi}$; 
the spectral index $n_{T}$ appearing in Eq. (\ref{EQ31a}) determines the high-frequency slope of the spectral energy density and it is given by:
\begin{equation}
 n_{T}(\epsilon, \delta) = \frac{1 - 3 \epsilon}{1- \epsilon} - \frac{\delta}{3 - \delta}, \qquad\qquad
n_{T}(w,\epsilon,\delta) = \frac{(1 -3\epsilon)}{(1 - \epsilon)} + \frac{9 (w -1) - (3 w-1)^2 \,\,\delta}{3 (3 w +1) - ( 3 w -1)^2 \,\,\delta}.
\label{EQ32}
\end{equation}
Both expressions of Eq. (\ref{EQ32}) coincide in the limit  $c_{st}\to 1$ and $w\to 1$.
 For $\delta \to 0$, $\epsilon \to 0$ and $w\to 1$ we have that  $n_{T} \to 1$ as originally discussed in \cite{TEN1}. 
 
In the plots of Fig. \ref{FIG1} the spectral index of Eq. (\ref{EQ32}) is illustrated for different 
values of $r_{T}$, $\delta$ and $w$. Since the current 
bounds on $r_{T}$ imply that $r_{T} < 0.06$ \cite{RT1,RT2}, according to the consistency relations (i.e. $r_{T} = 16\, \epsilon$) $\epsilon < 0.003$. In the right plot of Fig. \ref{FIG1} $r_{T}$ has been fixed at a value compatible with the current bounds while $\delta$ and $w$ are allowed to vary in their respective physical ranges.
\begin{figure}[!ht]
\centering
\includegraphics[height=5.9cm]{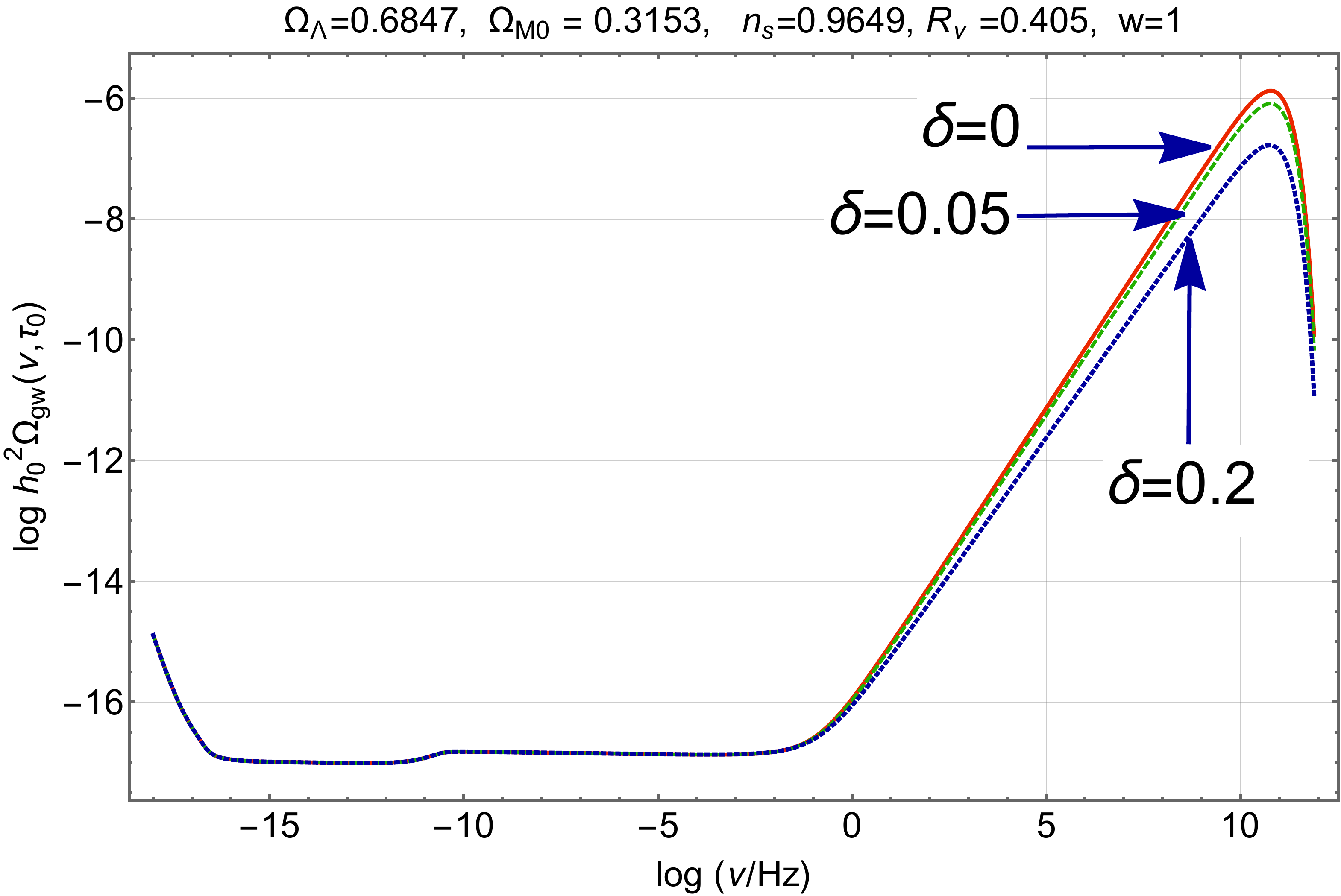}
\includegraphics[height=5.9cm]{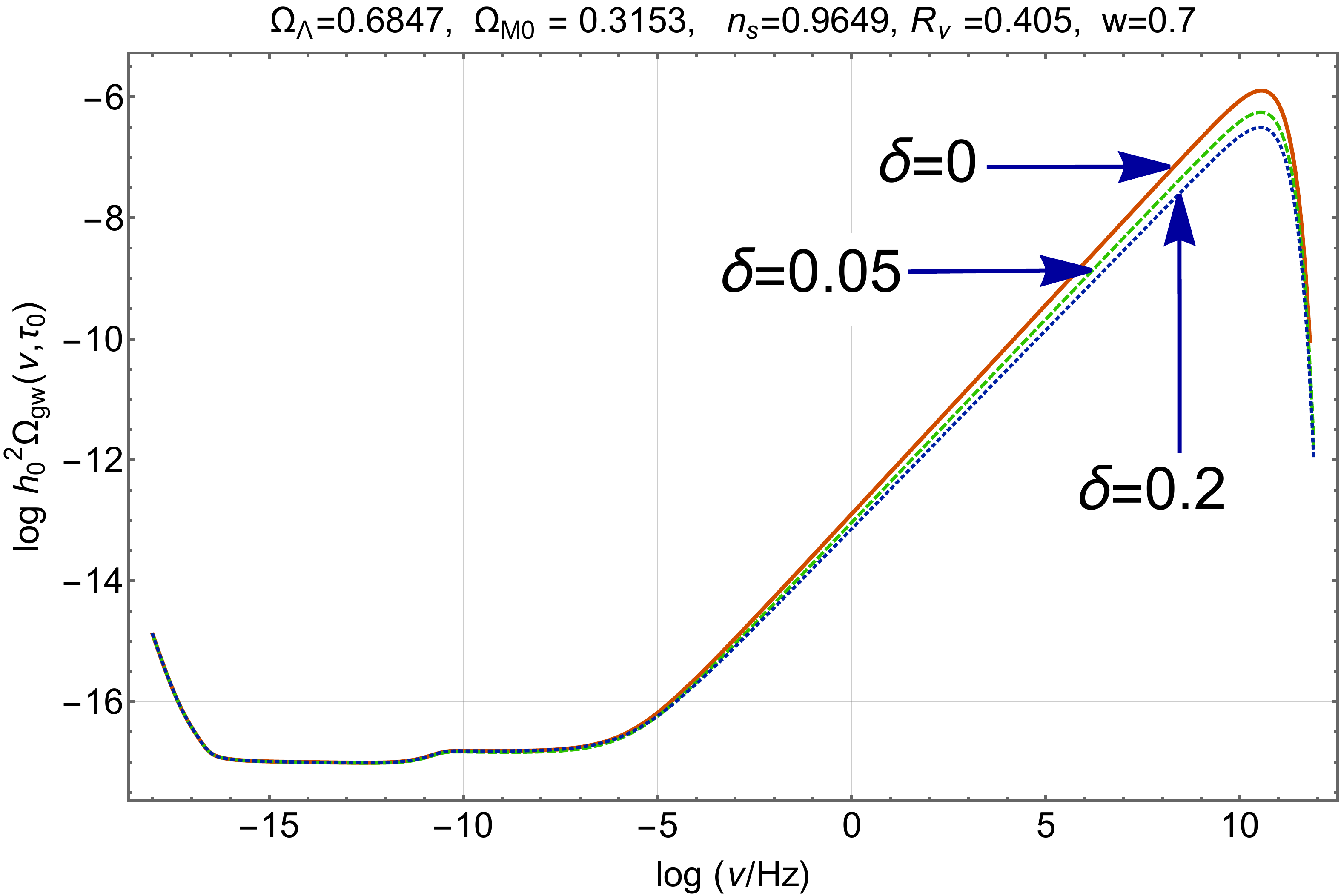}
\caption[a]{The spectral energy density of the relic gravitons is illustrated for different 
values of $\delta$ and $w$. We interpret this effect as an absorption of high-frequency gravitons by bulk viscous stresses. At lower frequencies the nearly scale-invariant plateau is suppressed by the free-streaming of the neutrinos which is controlled by $R_{\nu}$ (i.e. the neutrino fraction in the radiation plasma). Different values of $\Omega_{\Lambda}$ and $\Omega_{M0}$ 
only affect the overall normalization with an effect ${\mathcal O}(0.2)$.  We illustrated here the comoving frequencies that are related to the comoving wavenumbers as $\nu = k/(2\pi)$. Common logarithms are used on both axes.}
\label{FIG2}      
\end{figure}
The spectral index of Eq. (\ref{EQ32}) corresponds to the high-frequency slope of $\Omega_{gw}(\nu, \tau_{0})$
that has been computed numerically in Fig. \ref{FIG2} for a fiducial set of parameters \cite{RT1,RT2} and by employing 
the numerical approach described in Ref. \cite{TEN2}. While different sets of parameters do not crucially alter the results illustrated in Fig. \ref{FIG2}, we see from both plots that the high-frequency slope is modified, in practice, 
only if ${\mathcal O}(10^{-4})\leq \delta \ll 1$.  Depending on the value of $\delta$ the suppression can be of the same order of the damping due to the free-streaming of neutrinos.  Indeed, the quasi-flat plateau (corresponding to the wavelengths that leave the Hubble radius during inflation and reenter in the radiation stage) remains nearly unmodified except for the contribution of the neutrinos and for some possible outbreak of viscosity: between these two effects the former is more prominent than the latter. As Fig. \ref{FIG2} shows, below $100$ nHz the nearly scale-invariant plateau of inflationary origin suffers a $10$\% suppression due to the neutrino decoupling for typical temperatures ${\mathcal O}(\mathrm{MeV})$ \cite{FOUR,SEVEN}.  Around the quark-hadron phase transition a burst of  $\xi$ is compatible with the transverse momentum spectra and multiplicities of charged hadrons produced in the collisions of heavy ions at the energies of the Large Hadron Collider \cite{EIGHT} (see also \cite{NINE}). However, this effect does not seem sufficient to leave a clear imprint on the flat plateau of the relic gravitons and, in short, the reason is the following. In the radiation stage a sudden outbreak of the bulk viscosity implies that $a^{\prime\prime}/a \neq 0$ during a limited conformal time interval around the epoch of the phase transition. Since 
$a^{\prime\prime}/a$ also corresponds to the pump field of the tensor modes,
to modify the spectrum of the relic gravitons as the nHz frequencies cross the effective horizon we should have, at least, $\xi \, H/\rho_{t} = {\mathcal O}(1)$. This condition follows by requiring that
$a^{\prime\prime}/( a k^2) \simeq a^{\prime\prime}/( a \, {\mathcal H}^2) =  {\mathcal O}(1)$ since, at reentry, $k \simeq 1/\tau \sim {\mathcal H}$. Indeed, from Eq. (\ref{EQ12}) the magnitude of the required jump is estimated as:
\begin{equation}
\frac{a^{\prime\prime}}{a {\mathcal H}^2} =  \frac{3}{2} \frac{\xi \, \ell_{P}^2}{H} = \frac{{\mathcal O}(\mathcal{s} \, H)}{\rho_{t}} \leq {\mathcal O}(10^{-17}), \qquad\nu_{\xi}= 4.6 \biggl(\frac{{\mathcal N}}{10.75}\biggr)^{1/4} \biggl(\frac{T_{c}}{200\,\mathrm{MeV}}\biggr) 
\biggl(\frac{h_{0}^2 \Omega_{R0}}{4.15 \times 10^{-5}}\biggr)^{1/4}\,\,\mathrm{nHz},
\label{EQ33}
\end{equation}
where $\nu_{\xi}$ denotes the  comoving frequency that correspond to the Hubble radius at the time of the phase transition. According to (\ref{EQ33}) the burst in the bulk viscosity across the critical temperature (typically $T_{c} = {\mathcal O}(200)$ MeV) is, at most, of the order of $\xi_{0} \mathcal{s} H/\rho_{t}$. The value of $\xi_{0}$ can be directly 
taken from the analysis of Ref. \cite{EIGHT} and it is, at most, ${\mathcal O}(10)$. Thus, from Eq. (\ref{EQ33}) we will have that $\xi H/\rho_{t} = 10 \,{\mathcal O}(T_{c}/\overline{M}_{P})$. But since $T_{c}/\overline{M}_{P} = {\mathcal O}(10^{-20})$ the effect of the bulk viscosity amounts to a nearly invisible spike in the graviton spectrum for a typical frequency ${\mathcal O}(\mathrm{nHz})$. We assumed here that the entropy does not change too much across the phase transition but this effect marginally shifts the final estimate by one or two orders of magnitude from $10^{-19}$ to $10^{-17}$. 

The viscous absorption of high-frequency gravitons reentering during the radiation epoch is negligible except for the low-frequency modifications induced by neutrino free-streaming. However if the post-inflationary sound speed is stiffer than radiation  the spectral energy density can be modified in the ultra-high-frequency domain if the entropy density at onset of the post-inflationary stage is larger than $10^{-4}$ in units of $H_{1} \, \overline{M}_{P}^2$ where $H_{1}$ denotes the (nearly constant) inflationary expansion rate. In the extreme situation where the entropy density is ${\mathcal O}(0.1)\,\,H_{1} \, \overline{M}_{P}^2$ the suppression of the high-frequency spike is of the order of $10$\% and it is roughly comparable with the damping induced by the neutrino free-streaming in the nHz region. For the same frequency range an outbreak of bulk viscosity around the quark-hadron phase transition leads to a comparatively minute spike  in spectral energy density.  The present considerations suggest that the ultra-high-frequency gravitons, if ever detected, can be used to reconstruct not only the expansion rate prior to nucleosynthesis but also the viscous history of the post-inflationary medium.

I wish to thank T. Basaglia, A. Gentil-Beccot, S. Rohr and J. Vigen of the CERN Scientific Information Service for their usual kindness.

\end{document}